\documentclass[rnote]{aa} 

\usepackage{graphicx}
\usepackage{txfonts}

\begin{document}

\title{Chemically peculiar stars as seen with 2MASS}

\author{A.~Herdin\inst{1}
\and E.~Paunzen\inst{2}
\and M.~Netopil\inst{2}}
\institute{Universit{\"a}tssternwarte, T{\"u}rkenschanzstr. 17, A-1180 Wien, Austria
\email{epaunzen@physics.muni.cz}
\and
Department of Theoretical Physics and Astrophysics, Masaryk University,
Kotl\'a\v{r}sk\'a 2, CZ\,611\,37, Czech Republic}

   \date{} 
  \abstract
   {The chemically peculiar (CP) stars of the upper main sequence are well suited for investigating the 
          impact of magnetic fields and diffusion on the surface layers of slowly rotating stars. They can  
                even be traced in the Magellanic Clouds and are important to the understanding of the stellar formation and evolution.}
   {A systematic investigation of the near-infrared (NIR), 2MASS $JHK_{\rm s}$, photometry for the group of CP stars has never been performed.
          Nowadays, there is a great deal of data available in the NIR that  reach very large distances. It is therefore
          very important for CP stars to be unambiguously detected in the NIR region and for these detections to be used to derive astrophysical
          parameters (age and mass) by applying isochrone fitting. Furthermore, we investigated whether the CP stars behave in a  different way to  normal-type 
          stars in the various photometric diagrams.}
   {For our analysis, we carefully compiled a sample of CP and apparently normal (non-peculiar) type stars. Only stars
          for which high-quality (i.e. with low error levels), astrometric, and photometric data are available were chosen. In total,
                639 normal and 622 CP stars were selected and further analysed. All stars were dereddened and calibrated in 
                terms of the effective temperature and absolute magnitude (luminosity). Finally, isochrone fitting was applied.}
  %
         {No differences in the astrophysical parameters derived from 2MASS and Johnson $UBV$ photometry were found. Furthermore, 
          no statistical significant deviations from the normal type stars within several colour-colour and colour-magnitude diagrams
                were discovered. Therefore, it is not possible to detect new CP stars with the help of the photometric 2MASS colours only.
                A new effective temperature calibration, valid for all CP stars, using the $(V-K_{\mathrm S})_{\mathrm 0}$ colour was derived.}
   {}

\titlerunning{Chemically peculiar (CP) stars as seen with 2MASS}
\authorrunning{Herdin et al.}
\maketitle

\section{Introduction}

For many decades chemically peculiar (CP) stars have played an important role in stellar astrophysics. 
The reasons for this are mainly 
their  average slow rotation (even among the more massive stars on the main sequence), 
which makes them convenient stars for measuring rotational velocities; strong magnetic 
fields on the upper main sequence, used to understand the role of magnetic fields in 
the context of stellar evolution; their photometric variability; and their peculiar chemical composition
\citep{Miku10,Krti13}.
Chemically peculiar stars were divided into four groups by \citet{Pre74}: Am stars (CP1), 
Ap stars (CP2), HgMn stars (CP3), and He-weak stars (CP4). Observational progress unveiled other 
groups such as He-strong stars and $\lambda$ Bootis stars, and \citet{Smi96} created an 
extended classification scheme of CP stars. In addition CP2 stars can also be subdivided into 
several groups depending on their metal abundances (Eu, Sr, Sc, Cr, and so on). In this paper we 
follow the classification by Preston, but we also name the subgroups of the CP2 stars. 
The frequency of all CP stars in the solar neighbourhood on the main sequence ranges from 15 to 20\% 
for stars of spectral type B5--F0 \citep{Cat94,Rom07}.

In general, the CP stars seem to populate the whole age range from the zero-age main sequence (ZAMS) to 
the terminal-age main sequence (TAMS) which is in line with the currently used models \citep{Poe05}.

When determining fundamental parameters, photometry and spectroscopy are powerful tools to investigate 
these stars, though caution has to be taken when applying photometric calibrations to CP stars. The 
$uvby\beta$ luminosity calibration \citep{Cra79} leads to deviations in absolute magnitudes of CP2 
stars compared to the absolute magnitudes derived from Hipparcos data \citep{Mai00}. This effect 
occurs from peculiarities in the spectra of CP2 stars, which could lead to negative $E(b-y)$ 
values for many stars, therefore underestimating their reddening \citep{Fig98} such that additional calibration 
is recommended \citep{Ade79}. On the other hand, 
\citet{Net08} showed by means of cluster CP2 stars that there seems to be no significant 
influence on the reddening values. Unlike CP2 stars, the group of CP3 stars show no such peculiar 
behaviour and the $uvby\beta$ luminosity calibration can be applied \citep{Str66,Fig98}. 
The absolute magnitudes of CP1 stars are overestimated, hence a more accurate calibration of 
$M_{\rm V}$ has been developed \citep{Fig98} using the extension of the ordinary least squares estimator (BCES) 
method as described in \citet{Akr96} and the Hipparcos data. Helium-weak stars
have been calibrated in the BCD spectrophotometric system and the results are in good agreement 
with results from $UBV$ and Geneva photometry \citep{Cid07}. Helium-strong stars 
show different values when applying the BCD method and overestimate the effective temperature compared 
to the effective temperatures derived from $UBV$ and Geneva photometry \citep{Cid07}. 

However, when it comes to IR and near-IR (NIR)\ photometry, which are less sensitive to 
interstellar extinction than visual observations, very little is known about CP stars.
The only investigations in this respect were done by \citet{Gro83} and \citet{Kro87} who observed $JHKLM$ fluxes
for about 110 B- and A-type CP stars. \citet{Kro87} found no indications of dust shells or other circumstellar matter,
which should manifest as an IR-excess. In addition, they found no photometric variability with an amplitude larger 
than 0.05\,mag for their sample of stars. However, later on, variability in the NIR ($JHK$) with the same period, spectrum, and magnetic field variations as 
the optical light was detected, for example, for several stars of the SiSrCrEu subgroup
\citep{Cat98}. The amplitudes range from 0.04 to 0.002\,mag.

In this paper, we present the first extensive study of all four CP star subgroups using the photometric
data of the 2MASS survey \citep{Skr06}.

\begin{table}
\caption{The ZAMS \citep{Gir02} for solar metallicity (Z\,=\,0.019)
in terms of $M_{\mathrm K_{\mathrm S}}$ and $(V-K_{\mathrm S})_{\mathrm 0}$ for the investigated mass range
of B- to F-type stars.}
\label{ZAMS}
\begin{center}
\begin{tabular}{cccccc}
\hline
\hline 
$M_{\sun}$ & $M_{\mathrm K_{\mathrm S}}$ & $(V-K_{\mathrm S})_{\mathrm 0}$ & $M_{\sun}$ & $M_{\mathrm K_{\mathrm S}}$ & $(V-K_{\mathrm S})_{\mathrm 0}$ \\
           & [mag]                       & [mag]                           &            & [mag]                       & [mag]                           \\
\hline
1.3     &       +2.64   &       +1.081  &       4.0     &       +0.38   &       $-$0.409        \\
1.4     &       +2.40   &       +0.923  &       4.5     &       +0.16   &       $-$0.460        \\
1.5     &       +2.24   &       +0.752  &       5.0     &       $-$0.06 &       $-$0.502        \\
1.6     &       +2.11   &       +0.559  &       5.5     &       $-$0.27 &       $-$0.539        \\
1.7     &       +1.99   &       +0.385  &       6.0     &       $-$0.46 &       $-$0.573        \\
1.8     &       +1.90   &       +0.244  &       6.5     &       $-$0.64 &       $-$0.604        \\
1.9     &       +1.81   &       +0.142  &       7.0     &       $-$0.81 &       $-$0.632        \\
2.0     &       +1.70   &       +0.062  &       7.5     &       $-$0.98 &       $-$0.657        \\
2.1     &       +1.60   &       $-$0.002        &       8.0     &       $-$1.14 &       $-$0.680        \\
2.2     &       +1.51   &       $-$0.054        &       8.5     &       $-$1.32 &       $-$0.701        \\
2.3     &       +1.44   &       $-$0.095        &       9.0     &       $-$1.48 &       $-$0.720        \\
2.4     &       +1.38   &       $-$0.128        &       9.5     &       $-$1.64 &       $-$0.736        \\
2.5     &       +1.32   &       $-$0.157        &       10.0    &       $-$1.78 &       $-$0.750        \\
2.6     &       +1.25   &       $-$0.184        &       10.5    &       $-$1.92 &       $-$0.763        \\
2.7     &       +1.18   &       $-$0.209        &       11.0    &       $-$2.07 &       $-$0.774        \\
2.8     &       +1.11   &       $-$0.232        &       11.5    &       $-$2.23 &       $-$0.783        \\
2.9     &       +1.03   &       $-$0.253        &       12.0    &       $-$2.39 &       $-$0.792        \\
3.0     &       +0.96   &       $-$0.273        &       12.5    &       $-$2.54 &       $-$0.799        \\
3.5     &       +0.65   &       $-$0.350        &       13.0    &       $-$2.69 &       $-$0.805        \\
\hline
\end{tabular}
\end{center}
\end{table}

\begin{figure}
\begin{center}
\includegraphics[width=85mm]{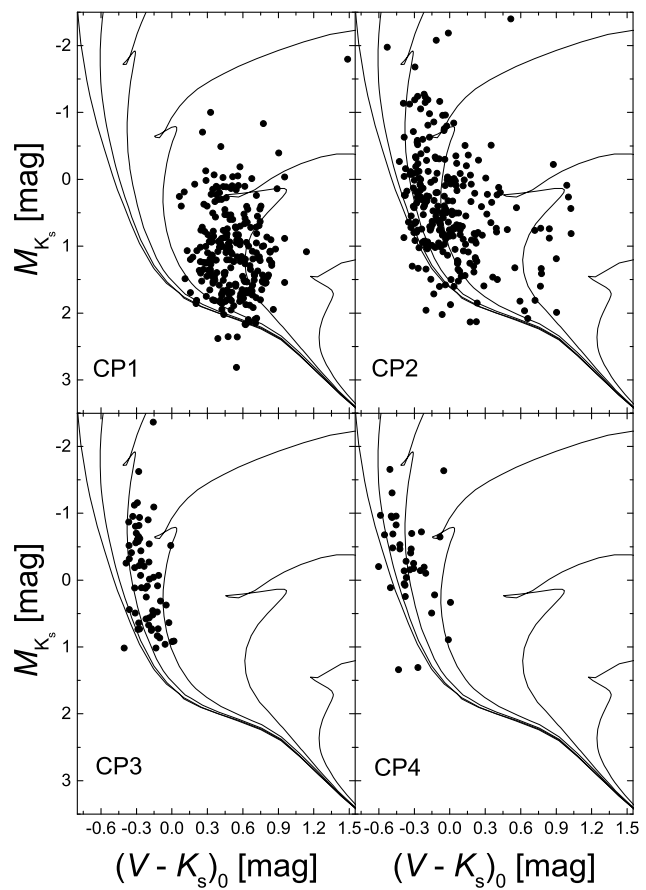}
\caption{The $M_{K_{\mathrm S}}$ versus $(V-K_{\mathrm S})_{\mathrm 0}$ diagrams for all CP subgroups together
with the isochrones from \citet{Gir02}.}
\label{isochrones}
\end{center}
\end{figure}

\section{Data sample}

For our analysis, we compiled a sample of CP and apparent normal-type or non-peculiar stars.

The original data sample consists of 956 CP stars of spectral types B0 to F9 (31\% B, 66\% A, 3\% F) 
from the General Catalogue of Ap and Am stars \citep{Ren09}. These objects are listed as best candidates
in this reference and are included within the new and improved reduction of the Hipparcos data catalogue 
from \citet{Lee07}.

Our original data sample of normal stars consists of 13\,637 main-sequence, luminosity class V, stars (8\% B, 30\% A, 62\% F). They were
chosen by the spectral classification listed by \citet{Gar94,Skif14}. In a series of four papers, \citet{Gar94}
investigated B- to F-type stars establishing new standard stars, which is especially valuable for our paper. 
As a first step, we removed all objects  listed as variable, 
peculiar, or in binary systems by cross-matching with the GCVS database 
\citep{Samu07}, the General Catalogue of Ap and Am stars \citep{Ren09}, and the Hipparcos catalogue \citep{Per97}.
Then, $uvby\beta$ photometry data were taken from \citet{Hau98}, parallaxes 
from Hipparcos data \citet{Lee07}, NIR photometry data from 2MASS \citep{Skr06}, and 
visual photometry data from the All-Sky Compiled Catalogue of 2.5 million stars \citep[ASCC,][]{Kha01} 
to produce a homogeneous data sample.

To obtain only high-quality data but still maintain a sufficiently large data sample, we restricted the 
error range of the data for parallaxes to $\sigma_{\rm \pi}$\,$/\pi$\,$\leq$\, 20\% and the 1-$\sigma$\ errors of the broadband photometric
measurements to $\sigma_{\rm \lambda}$\,$\leq$\ 0.1 mag for  $VJHK_{\rm s}$. The 1-$\sigma$\ errors of the narrowband? photometric indices were 
constrained as follows: $(b-y)\leq$ 0.027 mag, $m_{\rm 1}\leq$ 0.032 mag, and $c_{\rm 1}\leq$ 0.050 mag \citep{Bal94}. If the catalogues contained more than one 
measurement of a photometric index of a star, we used its mean value. Owing to insufficient $uvby\beta$ photometric data and the lack of accuracy, 
most of the normal stars were removed from the original sample. In addition to this removal, further reasons for the rejection due to the inaccuracy 
of other data sources are (from  most to least frequent) Hipparcos parallaxes, 
$uvby\beta$ dereddening, 2MASS photometry, and ASCC photometry. In addition, very few stars of the original sample were excluded for individual reasons, 
for example owing to an uncertain identification. Finally, 923 normal-type stars remain with high-quality data according to the above-listed standards.

Our data sample ranges in apparent magnitude and parallaxes from 4.439\,$\leq$\,$V$\,$\leq$\,10.237\,mag and
1.48\,$\leq$\,$\pi$\,$\leq$\,52.11\,mas for normal to 4.960\,$\leq$\,$V$\,$\leq$\,9.715\,mag and 
2.03\,$\leq$\,$\pi$\,$\leq$\,22.70\,mas for CP stars.

\section{Analysis}

The effective temperatures for all objects were obtained using the UVBYBETA calibration
by \citet{Napi93} with the improvement for CP stars published by \citet{Net08}. These calibrations
are well established and well tested.

We performed the dereddening for all stars also using  the calibration by \citet{Napi93}. 
Objects for which the dereddening did not converge after 500 iterations 
were excluded from  further analysis. These objects are mostly of luminosity class I, II, and III. To 
obtain the interstellar extinction-free apparent magnitudes in the NIR, we used a standard value for 
$E(B-V)$\,=\,1.35$ E(b-y)$ \citep{Cra75}. Stars with $E(b-y)$\,$<$\,0 (33.3\% of the sample, i.e. 165 normal and 237 CP stars) were treated as exhibiting 
no extinction,  and $E(b-y)$\ was set to zero. 
The final data sample was reduced to 639 normal stars of 
spectral type B1.5 to F9 (28\% B, 23\% A, and 49\% F) and 622 CP stars of 
spectral type B2 to F9 (32\% B, 64\% A, and 3\% F). The latter group consists of 263 CP1, 258 CP2, 63 CP3, and 38 CP4 stars.

The astrometric parallaxes of the Hipparcos catalogue \citep{Lee07} were transformed to absolute magnitudes ($M_{\mathrm K_{\mathrm S}}$) 
using the visual magnitude ($K_{\mathrm S}$) from \citet{Skr06} and the reddening ($A_{\mathrm K_{\mathrm S}}$\,=\,0.50$E(b-y)$) according to
$M_{\mathrm K_{\mathrm S}}$\,=\,5$\log\pi$\,+\,$K_{\mathrm S}$\,$-$\,$A_{\mathrm K_{\mathrm S}}$\,+\,5. The luminosities were 
calculated using the bolometric corrections   from \cite{Net08} scaled to $M_{\mathrm K_{\mathrm
S}}$ \citep{Mas06}.

Figure \ref{isochrones} shows the $M_{K_{\mathrm S}}$ versus $(V-K_{\mathrm S})_{\mathrm 0}$ diagrams for all CP subgroups together
with the isochrones from \citet{Gir02}. The latter range from 7.0\,$<$\,$\log t$\,$<$\,9.5\,dex, i.e. the full range 
of the main-sequence for B- to F-type stars. We chose this set of isochrones because it is well tested for a wide
variety of different star groups. However, we also used one set of isochrones from the newer PARSEC\footnote{http://stev.oapd.inaf.it/cgi-bin/cmd}
database. As expected, the differences are negligible. From the 622 CP stars only about 3\% are below the ZAMS.
This lends confidence not only in the use of the 2MASS colours, but also in the applied calibrations. In Table \ref{ZAMS},
the ZAMS for solar metallicity (Z\,=\,0.019) in terms of the mass $M_{\mathrm K_{\mathrm S}}$ and $(V-K_{\mathrm S})_{\mathrm 0}$ 
is listed. Our sample roughly covers the mass range from 1.3 to 9.0$M_{\sun}$. In the following, we discuss Fig. \ref{isochrones}
in more detail.
\begin{itemize}
\item CP1: Almost all objects are older than $\log t$\,$>$\,8.5\,dex with only a few objects close to the ZAMS.
This is consistent with the diffusion model that explains the observed chemical peculiarities. It is more efficient
in later stellar evolutionary stages and when  it has a longer time to work \citep{Mich83}.
\item CP2: The members of this group populate the whole age range from the ZAMS to about $\log t$\,=\,9.3\,dex.
This in line with the conclusions drawn by \citet{Poe05} who found that at least 16\% of the investigated CP2 
stars have fractional ages below 20\%.
\item CP3: This is the most homogeneous group in term of the age. Almost all the stars are located in the range
8.0\,$<$\,$\log t$\,$<$\,8.5\,dex. There are only a few stars at or close to the ZAMS.
\item CP4: There is wide range of ages from the ZAMS to about 8.5\,dex, which covers the complete stellar evolutionary
stages for the given mass range. 
\end{itemize}

The stellar fractional main-sequence ages ($t_{\rm frac}$) were derived using the effective temperature and luminosity by interpolating in 
the evolutionary tracks by 
\citet{Sch92} for solar metallicity (Z\,=\,0.020). This technique was already successfully applied to all CP subgroups
\citep{Wrai12,Pau13}. 
For the 2MASS system, the isochrones were taken from \citet{Gir02} for solar metallicity (Z\,=\,0.019) because the ones by
\citet{Sch92} are not available for the NIR region. Applying the above-mentioned method to the grids by \citet{Gir02} yielded compatible
results
for a representative sample of all subgroups within the errors of the effective temperatures and luminosities. For stars located outside the grids (below or above the main sequence), 
we adopted values for $t_{\rm frac}$ of either zero or 100\%.
However, more than 90\% of all CP stars are located well within the used evolutionary main-sequence borders. 
This supports the current knowledge that CP stars are in general main-sequence, luminosity class V, objects. We used the
$(VJHK_{\rm s})_{\rm 0}$ colours to get effective temperatures, luminosities, and fractional ages by simply using the 
smallest distance in  four-dimensional space within the solar abundant isochrones. Although this is not a very
accurate method, it is a good test if the 2MASS colours are useable for the calibration of CP stars. The comparison
with the values from the classical technique yielded the following median values of the differences for the complete
CP star sample: $\Delta T_\mathrm{eff}$\,=\,+90\,K, $\Delta \log L/L_\sun$\,=\,+0.01\,dex, and $\Delta t_{\rm frac}$\,=\,$-$5\%.
Splitting up the sample into the four subgroups did not result in any further improvements. From a statistical point of view,
the 2MASS colours can be used for CP stars to derive astrophysical parameters. This is especially important for studying
more distant and therefore fainter members of this group.

The distributions of the effective temperatures, luminosities, and calibrations of the effective temperature and luminosity of
the CP and normal-type objects were compared. We performed a Student's t-test \citep{Edig07} if both samples differed significantly  
from each other. On a 99.9\% significance level, the distributions of the two samples do not differ. Again, this supports the basic
assumption that the CP mechanism works from the ZAMS to the TAMS.

We inspected all colour-magnitude diagrams in the 2MASS system in order to search for a clear separation of CP to
normal stars. Such a discrimination is observed in the Johnson $UBV$ and Geneva seven-colour photometric systems \citep{Hau82}
mainly because most CP stars exhibit a UV flux deficiency and have several flux depressions such as the most prominent at 5200\,\AA\, 
\citep{Kod69,Lec73}. 
In addition, the reddening free $Q(JHK_{\rm S})$ parameter \citep{Com12}, which is defined as 
$Q(JHK_{\rm S})$\,=\,$(J-H)$\,+\,$\alpha(H-K_{\mathrm S})$ with $\alpha$\,=\,$\frac{E(J-H)}{E(H-K_{\mathrm S})}$, was investigated.
The actual values of $\alpha$ range between 1.4 and 2.0 \citep{Yua13}. No statistical significant deviation from the normal-type stars was discovered. Therefore, it is not possible to detect new CP stars with the help of the 2MASS colours only. 
\citet{Leo91} compared several $UBVJH$ colour-colour diagrams of normal and CP2 stars and came to a similar
conclusion.

\begin{figure}
\begin{center}
\includegraphics[width=85mm]{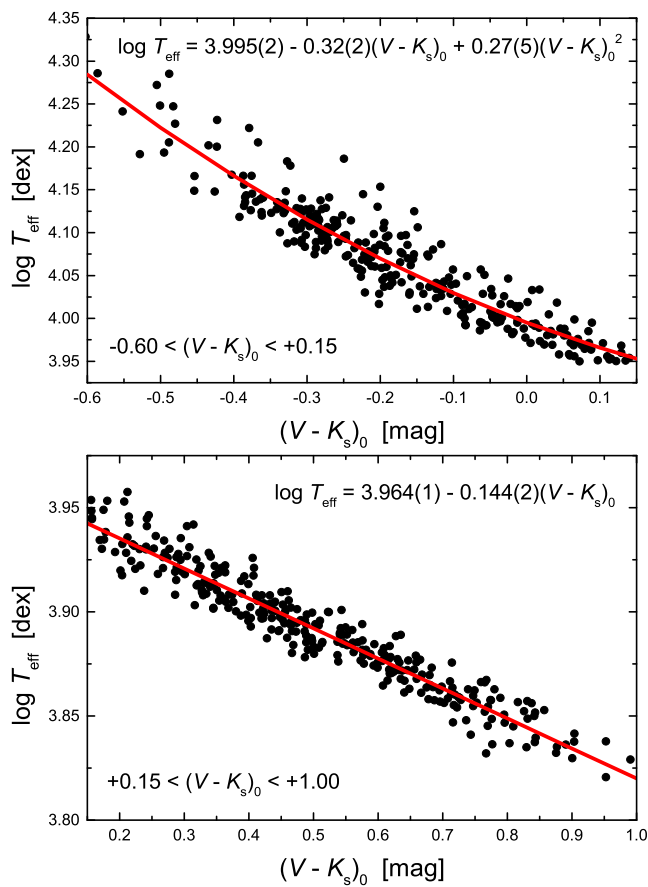}
\caption{The effective temperature calibration for all four CP star groups in terms of $(V-K_{\mathrm S})_{\mathrm 0}$.}
\label{calibration}
\end{center}
\end{figure}

As the final step, the effective temperatures derived from the $uvby\beta$ photometric data were plotted versus the 
$(V-K_{\mathrm S})_{\mathrm 0}$ colours in order to derive a temperature calibration for CP stars. There are only
a few stars that significantly deviate from the relation shown in Fig. \ref{calibration}. These objects are, with a few
exceptions, close or spectroscopic binary systems of the CP1 subgroup and were discarded from the analysis. We fitted a line
for the cooler stars and a polynomial of first order to the hotter stars for the complete sample not divided into subgroups
(Fig. \ref{calibration}). The derived relations and their valid colour ranges are
\begin{itemize}
\item $\log T_\mathrm{eff}$\,=\,$3.995(2)\,$-$\,0.32(2)(V-K_{\mathrm S})_{\mathrm 0}$+0.27(5)$(V-K_{\mathrm S})_{\mathrm 0}^{2}$ for $-$0.60\,$<$\,$(V-K_{\mathrm S})_{\mathrm 0}$\,$<$\,+0.15\,mag; $N$\,=\,294
\item $\log T_\mathrm{eff}$\,=\,3.964(1)\,$-$\,0.144(2)$(V-K_{\mathrm S})_{\mathrm 0}$ for +0.15\,$<$\,$(V-K_{\mathrm S})_{\mathrm 0}$\,$<$\,+1.00\,mag; $N$\,=\,301
\end{itemize}
These calibrations cover an effective temperature range from 6600 to 19\,000\,K. 
Taking the limiting colour magnitudes for both calibrations, we find a maximum error of $\pm$13\% and $\pm$3\%, respectively.
Merging both subsamples and performing a polynomial fit up to the third order did not result in a smaller error. 

\section{Conclusions}

Current estimates suggest that about 15\% of all B- to F-type stars belong to
CP groups. To date, more than 8\,000 probable CP objects (or candidates) are known in the Milky Way
\citep{Ren09}, of which nearly one-half belong to CP1 (i.e. non-magnetic) and CP2/4 (i.e. magnetic) stars. The
group of CP3 (HgMn) objects play a minor part with only about 200 candidates listed in the aforementioned
reference.

The group of CP stars is an excellent laboratory with  which to test astrophysical mechanisms such as   diffusion (both microscopic and
turbulent), convection, meridional circulation, mass loss, and accretion in the absence/presence
of an organized stellar magnetic field.

For the first time we investigated  statistically the behaviour of the CP stars in the NIR
using the 2MASS $JHK_{\rm s}$ colours together with the Johnson $V$ data. A sample of CP and normal-type
stars was selected on the basis of available Str{\"o}mgren $uvby\beta$ colours and accurate Hipparcos
parallaxes. 

All stars were dereddened and calibrated in terms of the effective temperature and absolute magnitude (luminosity)
using well-tested calibrations. A new effective temperature calibration for CP stars using the 
$(V-K_{\mathrm S})_{\mathrm 0}$ colour was derived.

We conclude that the 2MASS colours are well suited to statistically deriving the astrophysical parameters  of CP stars, but do not allow to  distinguish them a priori from normal-type objects. 
However, one has to keep in mind that several subgroups of these stars are
variable over the whole spectral range, which introduces an additional error source for the calibration process.
This result is important for the
study of more distant and therefore faint CP candidates for which no  photometric data other than those from  Johnson $V$ and
2MASS  are available. 

\begin{acknowledgements}
This project is financed by the SoMoPro II programme (3SGA5916). The research leading
to these results has acquired a financial grant from the People Programme
(Marie Curie action) of the Seventh Framework Programme of EU according to the REA Grant
Agreement No. 291782. The research is further co-financed by the South-Moravian Region. 
It was also supported by the grants GP14-26115P, 7AMB14AT015, and
the financial contributions of the Austrian Agency for International 
Cooperation in Education and Research (BG-03/2013 and CZ-09/2014).
This work reflects only the author's views and the European 
Union is not liable for any use that may be made of the information contained therein.
\end{acknowledgements}


\begin{thebibliography}{}
\bibitem[\protect\citeauthoryear{Adelman}{1979}]{Ade79} Adelman, S. J. 1979, \aj, 84, 857
\bibitem[\protect\citeauthoryear{Akritas \& Bershady}{1996}]{Akr96} Akritas, M. G., \& Bershady, M. A. 1996, \apj, 470, 706
\bibitem[\protect\citeauthoryear{Balona}{1994}]{Bal94} Balona, L. A. 1994, \mnras, 268, 119
\bibitem[\protect\citeauthoryear{Catalano \& Leone}{1994}]{Cat94} Catalano, F. A., \& Leone, F. 1994, in The MK Process at 50 Years: a powerful
tool for astrophysical insight, eds. C.J. Corbally, R.O. Gray, \& R.F. Garrison, ASP Conf. Ser., 60, 110
\bibitem[\protect\citeauthoryear{Catalano et al.}{1998}]{Cat98} Catalano, F. A., Leone, F., \& Kroll, R. 1998, \aaps, 131, 63
\bibitem[\protect\citeauthoryear{Cidale et al.}{2007}]{Cid07} Cidale, L. S., Arias, M. L., Torres, A. F., et al. 2007, \aap, 468, 263
\bibitem[\protect\citeauthoryear{Comer{\' o}n \& Pasquali}{2012}]{Com12} Comer{\' o}n, F., \& Pasquali, A. 2012, \aap, 543, A101
\bibitem[\protect\citeauthoryear{Crawford}{1975}]{Cra75} Crawford, D. L. 1975, \aj, 80, 955
\bibitem[\protect\citeauthoryear{Crawford}{1979}]{Cra79} Crawford, D. L. 1979, \aj, 84, 1858
\bibitem[\protect\citeauthoryear{Edgington \& Onghena}{2007}]{Edig07} Edgington, E. S., \& Onghena, P. 2007, Randomization Tests, Fourth Edition (Chapman and Hall/CRC, London)
\bibitem[\protect\citeauthoryear{Figueras et al.}{1998}]{Fig98} Figueras, F., Luri, X., Gomez, A. E.,  et al. 1998, Contributions of the Astronomical 
Observatory Skalnate Pleso, 27, 184
\bibitem[\protect\citeauthoryear{Girardi et al.}{2002}]{Gir02} Girardi, L., Bertelli, G., Bressan, A., et al. 2002, \aap, 391, 195
\bibitem[\protect\citeauthoryear{Garrison \& Gray}{1994}]{Gar94} Garrison, R. F., \& Gray, R. O. 1994, \aj, 107, 1556
\bibitem[\protect\citeauthoryear{Groote \& Kaufmann}{1983}]{Gro83} Groote, D., \& Kaufmann, J. P. 1983, \aaps, 53, 91
\bibitem[\protect\citeauthoryear{Hauck \& Mermilliod}{1998}]{Hau98} Hauck, B., \& Mermilliod, M. 1998, \aaps, 129, 431 
\bibitem[\protect\citeauthoryear{Hauck \& North}{1982}]{Hau82} Hauck, B., \& North, P. 1982, \aap, 114, 23
\bibitem[\protect\citeauthoryear{Kharchenko}{2001}]{Kha01} Kharchenko, N. V. 2001, Kinematika i Fizika Nebesnykh Tel, vol. 17, 409
\bibitem[\protect\citeauthoryear{Kodaira}{1969}]{Kod69} Kodaira, K. 1969, \apj, 157, 59
\bibitem[\protect\citeauthoryear{Kroll et al.}{1987}]{Kro87} Kroll, R., Schneider, H., Voigt, H. H., \& Catalano, F. A. 1987, \aaps, 67, 195
\bibitem[\protect\citeauthoryear{Krti\v{c}ka et al.}{2013}]{Krti13} Krti\v{c}ka, J., Jan{\' i}k, J., Markov{\' a}, H. et al. 2013, \aap, 556, A18
\bibitem[\protect\citeauthoryear{Leckrone}{1973}]{Lec73} Leckrone, D. S. 1973, \apj, 185, 577
\bibitem[\protect\citeauthoryear{Leone \& Catalano}{1991}]{Leo91} Leone, F., \& Catalano, F. A. 1991, \aap, 242, 199
\bibitem[\protect\citeauthoryear{Maitzen et al.}{2000}]{Mai00} Maitzen, H. M., Paunzen, E., Vogt, N., \& Weiss, W. W. 2000, \aaps, 355, 1003
\bibitem[\protect\citeauthoryear{Masana et al.}{2006}]{Mas06} Masana, E., Jordi, C., \& Ribas, I. 2006, \aap, 450, 735
\bibitem[\protect\citeauthoryear{Michaud \& Charland}{1986}]{Mich86} Michaud, G., \& Charland, Y. 1986, \apj, 311, 326
\bibitem[\protect\citeauthoryear{Michaud et al.}{1983}]{Mich83} Michaud, G., Tarasick, D., Charland, Y., \& Pelletier, C. 1983, \apj, 269, 239
\bibitem[\protect\citeauthoryear{Mikul{\'a}\v{s}ek et al.}{2010}]{Miku10} Mikul{\'a}\v{s}ek, Z., Krti\v{c}ka, J., 
Henry, G. W., de Villiers, S. N., Paunzen, E., \& Zejda, M. 2010, \aap, 511, L7
\bibitem[\protect\citeauthoryear{Napiwotzki et al.}{1993}]{Napi93} Napiwotzki, R., Schoenberner, D., \& Wenske, V. 1993, \aap, 268, 653 
\bibitem[\protect\citeauthoryear{Netopil et al.}{2008}]{Net08} Netopil, M., Paunzen, E., Maitzen, H. M., North, P., \& Hubrig, S. 2008, \aap, 491, 545 
\bibitem[\protect\citeauthoryear{Paunzen et al.}{2013}]{Pau13} Paunzen, E., Wraight, K. T., Fossati, L., et al. 2013, \mnras, 429, 119
\bibitem[\protect\citeauthoryear{Perryman et al.}{1997}]{Per97} Perryman, M. A. C., Lindegren, L., Kovalevsky, J., et al. 1997, \aap, 323, L49
\bibitem[\protect\citeauthoryear{P{\"o}hnl et al.}{2005}]{Poe05} P{\"o}hnl, H., Paunzen, E., \& Maitzen, H. M. 2005, \aap, 441, 1111
\bibitem[\protect\citeauthoryear{Preston}{1974}]{Pre74} Preston, G. W. 1974, \araa, 12, 257
\bibitem[\protect\citeauthoryear{Renson \& Manfroid}{2009}]{Ren09} Renson, P., \& Manfroid, J. 2009, \aap, 498, 961
\bibitem[\protect\citeauthoryear{Romanyuk}{2007}]{Rom07} Romanyuk, I. I. 2007, Astrophysical Bulletin 62, 62
\bibitem[\protect\citeauthoryear{Samus et al.}{2007 -- 2014}]{Samu07} Samus, N. N., Durlevich, O. V., Kazarovets, E. V., 
et al. 2007 -- 2014, General Catalogue of 
Variable Stars, VizieR On-line Catalog (http://cdsarc.u-strasbg.fr/viz-bin/Cat?B/gcvs)
\bibitem[\protect\citeauthoryear{Schaller et al.}{1992}]{Sch92} Schaller, G., Schaerer, D., Meynet, G., \& Maeder, A. 1992, \aaps, 96, 269 
\bibitem[\protect\citeauthoryear{Skrutskie et al.}{2006}]{Skr06} Skrutskie, M. F., Cutri, R. M., Stiening, R., et al. 2006, \aj, 131, 1163
\bibitem[\protect\citeauthoryear{Skiff}{2014}]{Skif14} Skiff, B. A. 2014, Catalogue of Stellar Spectral Classifications, VizieR Online Data Catalog
(http://cdsarc.u-strasbg.fr/viz-bin/VizieR?-source=B/mk)
\bibitem[\protect\citeauthoryear{Smith}{1996}]{Smi96} Smith, K. C. 1996, \apss, 237, 77
\bibitem[\protect\citeauthoryear{Strai\u{z}ys \&  Lazauskait\.{e}}{2009}]{Str09} Strai\u{z}ys, V., \&  Lazauskait\.{e} R., 2009, Balt. Astron., 18, 19
\bibitem[\protect\citeauthoryear{Str{\"o}mgren}{1966}]{Str66} Str{\"o}mgren, B. 1966, \araa, 4, 433
\bibitem[\protect\citeauthoryear{van Leeuwen}{2007}]{Lee07} van Leeuwen, F. 2007, \aap, 474, 653
\bibitem[\protect\citeauthoryear{Yuan et al.}{2013}]{Yua13} Yuan, H. B., Liu, X. W., \& Xiang, M. S. 2013, \mnras, 430, 2188
\bibitem[\protect\citeauthoryear{Wraight et al.}{2012}]{Wrai12} Wraight, K. T., Fossati, L., Netopil, M., et al. 2012, \mnras, 420, 757
\end{thebibliography}
\end{document}